\documentclass[trackchanges,twocolumn]{aastex701}

\usepackage{amsmath}	
\begin{document}
\title{High-Frequency Transverse Waves in Chromospheric Fibrils Revealed by DKIST}

\author{Qifan Dong} 
\affiliation{Yunnan Observatories, Chinese Academy of Sciences, Kunming 650216, People's Republic of China}
\affiliation{University of Chinese Academy of Sciences, Beijing 100049, People's Republic of China}
\email{dongqifan@ynao.ac.cn}

\author[0000-0003-2891-6267]{Xiaoli Yan}
\affiliation{Yunnan Observatories, Chinese Academy of Sciences, Kunming 650216, People's Republic of China}
\affiliation{Yunnan Key Laboratory of Solar Physics and Space Science, Kunming 650216, People's Republic of China}
\email{yanxl@ynao.ac.cn}
\correspondingauthor{Xiaoli Yan}
\email{yanxl@ynao.ac.cn}

\author[0000-0002-6526-5363]{Zhike Xue}
\affiliation{Yunnan Observatories, Chinese Academy of Sciences, Kunming 650216, People's Republic of China}
\affiliation{Yunnan Key Laboratory of Solar Physics and Space Science, Kunming 650216, People's Republic of China}
\email{zkxue@ynao.ac.cn}

\author[0000-0003-0236-2243]{Liheng Yang}
\affiliation{Yunnan Observatories, Chinese Academy of Sciences, Kunming 650216, People's Republic of China}
\affiliation{Yunnan Key Laboratory of Solar Physics and Space Science, Kunming 650216, People's Republic of China}
\email{yangliheng@ynao.ac.cn}

\author[0000-0003-4393-9731]{Jincheng Wang}
\affiliation{Yunnan Observatories, Chinese Academy of Sciences, Kunming 650216, People's Republic of China}
\affiliation{Yunnan Key Laboratory of Solar Physics and Space Science, Kunming 650216, People's Republic of China}
\email{wangjincheng@ynao.ac.cn}

\author{Yadan Duan}
\affiliation{Yunnan Observatories, Chinese Academy of Sciences, Kunming 650216, People's Republic of China}
\affiliation{Yunnan Key Laboratory of Solar Physics and Space Science, Kunming 650216, People's Republic of China}
\email{duanyadan@ynao.ac.cn}

\author[0000-0002-9121-9686]{Zhe Xu}
\affiliation{Yunnan Observatories, Chinese Academy of Sciences, Kunming 650216, People's Republic of China}
\affiliation{Yunnan Key Laboratory of Solar Physics and Space Science, Kunming 650216, People's Republic of China}
\email{xuzhe6249@ynao.ac.cn}


\begin{abstract}

Chromospheric fibrils are ubiquitous dynamical structures that can support magnetohydrodynamic waves and may contribute to energy transport in the solar atmosphere. Using high-resolution observations from the Visible Broadband Imager (VBI) onboard the Daniel K. Inouye Solar Telescope (DKIST), we investigate transverse oscillations of H$\alpha$ fibrils. Our observations encompass both quiet-Sun and active regions. The results reveal that high-frequency transverse oscillations are highly prevalent in fibrils across different magnetic environments. The detected oscillations exhibit typical periods of 56 $\pm$ 49 s, displacement amplitudes of 37 $\pm$ 25 km, velocity amplitudes of 6 $\pm$ 4 km\,s$^{-1}$, and phase velocities of 21 $\pm$ 13 km\,s$^{-1}$. Under typical chromospheric conditions, the estimated wave energy flux ranges from 20 to 200 W\,m$^{-2}$. The characteristics of these transverse oscillations are consistent with those of Alfv\'{e}nic kink-mode waves. Our results reveal the existence of the high-frequency, small-amplitude transverse waves in chromospheric fibrils and providing new constraints on their contribution to the chromospheric and coronal heating.

\end{abstract}


\section{Introduction} \label{sec:intro}

Solar chromosphere contains a wealth of fine-scale structures, including spicules, mottles, fibrils, and rapid blue/red shifted excursions (RBEs/RREs), which are believed to play important roles in the transport of mass and energy throughout the solar atmosphere (e.g., \citealp{DePontieu+etal+2007pasj,Rutten+2019,Sterling+2021,Skirvin+2023}). Among these features, fibrils are ubiquitous elongated structures observed in chromospheric spectral lines such as H$\alpha$ and \ion{Ca}{2}, appearing as thin, thread-like structures that outline the magnetic field lines \citep{Leenaarts+2012,Jafarzadeh+2017+APJS}. In the quiet-Sun (QS), fibrils are often associated with magnetic bright points and internetwork magnetic fields, whereas in active region (AR), they form more organized patterns around sunspots and plage regions (e.g., \citealp{Leenaarts+2015}). Owing to their abundance, long lifetimes, and well-defined morphology, fibrils provide ideal waveguides for investigating magnetohydrodynamic (MHD) wave propagation and energy transport in the chromosphere. Consequently, transverse oscillations of fibrils have become a key diagnostic for probing chromospheric wave dynamics and energy transport (e.g., \citealp{Druett+2022}).

Early observations of transverse motions in the chromosphere were largely obtained from studies of spicules (e.g., \citealp{Pasachoff+1968,DePontieu+2007Science,Zaqarashvili+2009SSRv,Shetye+2021}). Similar transverse motions were subsequently identified in other chromospheric structures, including fibrils (e.g., \citealp{Morton+2012+NC,Jafarzadeh+2017+osc}), mottles (e.g., \citealp{Kuridze+2012,Kuridze+2013}), and RBEs (e.g., \citealp{Sekse+2013}). These transverse motions were initially interpreted as Alfv{\'e}n waves (e.g., \citealp{Tomczyk+2007Science,He+2009,Mathioudakis+2013}), although some studies suggested that they may instead represent kink waves (e.g., \citealp{VanDoorsselaere+2008,VanDoorsselaere+2014,Mooroogen+2017}). However, subsequent theoretical studies showed that in magnetic flux tubes, such transverse motions are more appropriately described as kink-mode waves with Alfv{\'e}nic properties (e.g., \citealp{Kwak+2023}, see also the reviews \citealp{Jess+2023LRSP,Morton+2023RvMP}). Specifically, they occur in inhomogeneous plasmas and exhibit key features of Alfv{\'e}n waves, including incompressibility and magnetic tension as the primary restoring force (e.g., \citealp{Alfven+1942,Jess+2015SSRv}). These waves are considered promising carriers of energy transport, capable of transferring energy into the upper solar atmosphere and potentially contributing to chromospheric and coronal heating (e.g., \citealp{Kumar+2017NatSR,VanDoorsselaere+2020SSRv,Lim+2023}).

One of the primary motivations for studying transverse oscillations is to determine whether the associated wave energy flux is sufficient to contribute to chromospheric heating. However, previous estimates of the energy flux vary widely, largely due to differences in observational instruments, target structures, and analysis methods. Observational evidence for high-frequency transverse motions has been particularly abundant in spicules \citep{Bate+2022}. High-cadence observations from Hinode and the Rapid Oscillations in the Solar Atmosphere (ROSA) have revealed ubiquitous short-period transverse motions in spicules, with periods ranging from several seconds to approximately one minute \citep{Okamoto+2011ApJ,Bate+2022}. These observations suggest that spicules may provide an important pathway for transporting high-frequency Alfv\'{e}nic energy into the upper solar atmosphere. Similar high-frequency transverse motions have also been identified in chromospheric fibrils. Using SUNRISE/SuFI observations, \citet{Jafarzadeh+2017+osc} reported ubiquitous transverse oscillations in slender \ion{Ca}{2} H fibrils, with median velocity amplitudes and periods of 2.4 $\pm$ 0.8 km\,s$^{-1}$ and 83 $\pm$ 29 s, respectively. More recently, \citet{Petrova+2025} detected a short-period transverse oscillation in a fibril observed with the MiHI prototype, characterized by a period of 15 s and a displacement amplitude of 42 km, further demonstrating the presence of high-frequency oscillations in fibrils.

However, previous studies have generally been limited by sample size, or instrumental resolution, leaving the prevalence and statistical properties of high-frequency transverse waves across different chromospheric environments insufficiently characterized. The Daniel K. Inouye Solar Telescope (DKIST; \citealp{Rimmele+2020}), with its unprecedented spatial and temporal resolution, provides new opportunities to investigate chromospheric wave dynamics at previously unresolved spatial and temporal scales. In particular, observations from the Visible Broadband Imager (VBI; \citealp{Woger+2021}) enable the detection of subtle transverse motions in individual fibrils that are easily obscured in previous observations. In this work, H$\alpha$ observations obtained with the DKIST/VBI are used to investigate transverse oscillations of chromospheric fibrils in both QS and AR.

The remainder of this paper is organized as follows. Section \ref{sec:OaM} describes the DKIST/VBI observations and the methodology used to identify and track transverse oscillations. Section \ref{sec:results} presents the observational results and statistical analyses. Section \ref{sec:dis} discusses the results and summarizes the main conclusions.

\begin{figure*}[ht!]
	\includegraphics[scale=1.0]{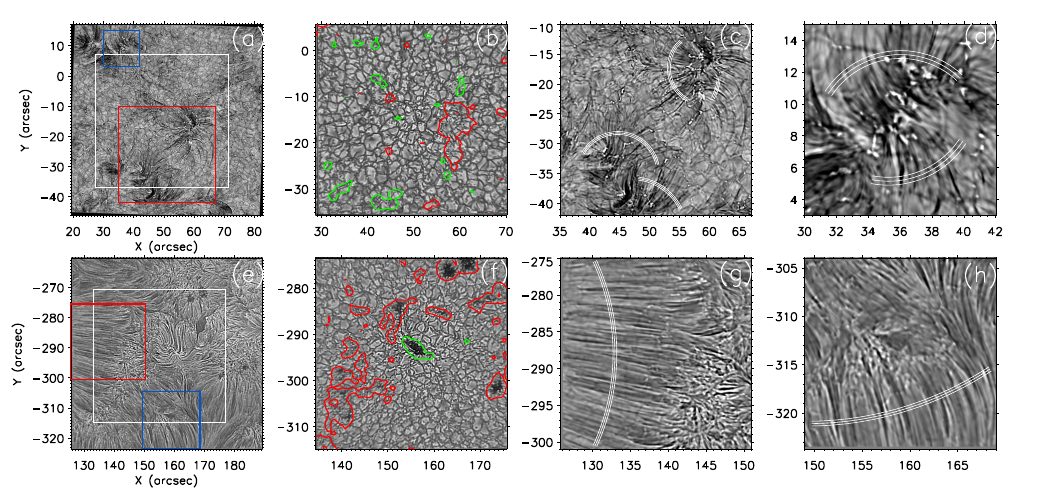}
	\caption{Overview of the QS and AR observations. Panels (a) and (e) show the H$\alpha$ images of the QS and AR, respectively. Panels (b) and (f) show the corresponding 450.3 nm blue continuum images, with fields of view indicated by the white boxes in panels (a) and (e). The red and green contours in panels (b) and (f) represent positive and negative polarity SDO/HMI LOS magnetic fields, respectively, at levels of $\pm$ 20 G for the QS and $\pm$ 300 G for the AR. Panels (c) and (g) show enlarged views of the regions outlined by the red boxes in panels (a) and (e), while panels (d) and (h) show enlarged views of the regions outlined by the blue boxes. The slit locations used for the time–distance analysis are marked by white lines.}
	\label{fig:fig1}
\end{figure*}

\section{Observation and Method} \label{sec:OaM}
\subsection{Observation} \label{subsec:Ob}

To investigate the transverse oscillations of chromospheric fibrils, this study employs high spatio-temporal resolution dual-channel coordinated observations obtained with the VBI. As one of the first-light instruments, the VBI provides simultaneous red and blue broadband channels designed to deliver high-resolution observations of the photosphere and chromosphere at the diffraction limit of the telescope \citep{Woger+2021}, providing excellent conditions for tracking fine-scale plasma evolution in the solar atmosphere.
 
The primary data analyzed in this study were acquired with the VBI red channel (550–860 nm) in the H$\alpha$ line centered at 656.289 nm, with a full width at half maximum (FWHM) of 0.047 nm. This passband offers outstanding sensitivity for detecting fine chromospheric structures such as fibrils (e.g., \citealp{Pereira+2016}). The spatial sampling of this channel is 0\farcs017 pixel$^{-1}$, with a field of view (FOV) of approximately $61^{\prime\prime} \times 61^{\prime\prime}$. In addition, to provide the morphological context of the underlying photosphere, co-temporal blue continuum images from the VBI blue channel at 450.3 nm were utilized. This channel (FWHM = 0.406 nm) provides a spatial sampling of 0\farcs011 pixel$^{-1}$. The two channels are highly synchronized during the observation, with short-exposure raw frames continuously acquired within a narrow temporal window \citep{Woger+2021}. These frames are subsequently reconstructed using adaptive optics correction combined with post-facto speckle reconstruction techniques, resulting in a final temporal cadence of 2.667 s \citep{Woger+2008,Woger+2021}. The VBI provides both high spatial resolution and the capability to capture rapid transverse oscillations, enabling the detection of small-amplitude and high-frequency oscillations (periods shorter than 50 s).To further characterize the underlying magnetic environment, line-of-sight (LOS) magnetograms from the Helioseismic and Magnetic Imager (HMI) aboard the Solar Dynamics Observatory (SDO; \citealp{Pesnell+2012SoPh}) were utilized. HMI provides full-disk LOS magnetic field measurements with a spatial sampling of 0\farcs5 pixel$^{-1}$ and a cadence of 45 s \citep{Schou+2012SoPh,Scherrer+2012SoPh}. The LOS magnetograms were spatially co-aligned with the VBI 450.3 nm continuum images, and magnetic field contours were overplotted to identify the distribution and polarity of the photospheric magnetic field.

To compare the transverse oscillation properties of fibrils under different magnetic environments, we analyzed two independent data sets obtained from the DKIST Data Center Archive, corresponding to the QS and the AR, respectively.

The first data set (D1) is the observation of a QS region located near disk center, centered at (50\farcs9, -14\farcs5) (Figure \ref{fig:fig1}(a)). D1 was obtained from 20:28:13 to 20:59:04 UT on 2023 August 29 (Product ID: L1-TDWAC), with a Fried parameter (r\textsubscript{0}) of 11.7 cm. The corresponding 450.3 nm continuum images show slowly evolving granulation with only a few magnetic bright points. The LOS magnetic field contours are overlaid at $\pm$ 20 G (Figure \ref{fig:fig1}(b)).

The second data set (D2) is the observation of an AR centered at (155\farcs1, -293\farcs2) (Figure \ref{fig:fig1}(e)). D2 was obtained from 20:09:45 to 20:46:06 UT on 2024 August 27 (Product ID: L1-FTNAW), with an r\textsubscript{0} of 7 cm. In contrast to D1, the 450.3 nm continuum images show clear photospheric signatures of strong magnetic activity, including pores and magnetic bright points. For D2, the LOS magnetic field contours are overlaid at $\pm$ 300 G (Figure \ref{fig:fig1}(f)).

\subsection{Method} \label{subsec:method}

Accurate tracking of transverse oscillations requires enhanced visibility of the fine fibril structures. Multiscale Gaussian Normalization (MGN; \citealp{Morgan+2014}) was therefore applied to each individual frame. This method enhances local contrast by normalizing intensity variations relative to the local mean and standard deviation calculated at multiple spatial scales using Gaussian convolution. For the VBI data, a customized set of relatively small Gaussian kernels was adopted, with widths of $\omega$ = 0.8, 1.6, 3.2, 6, 12, 24 pixels. This choice avoids over-enhancing large-scale background structures and improves the visibility of fine chromospheric fibrils.

\begin{figure}[ht!]
	\plotone{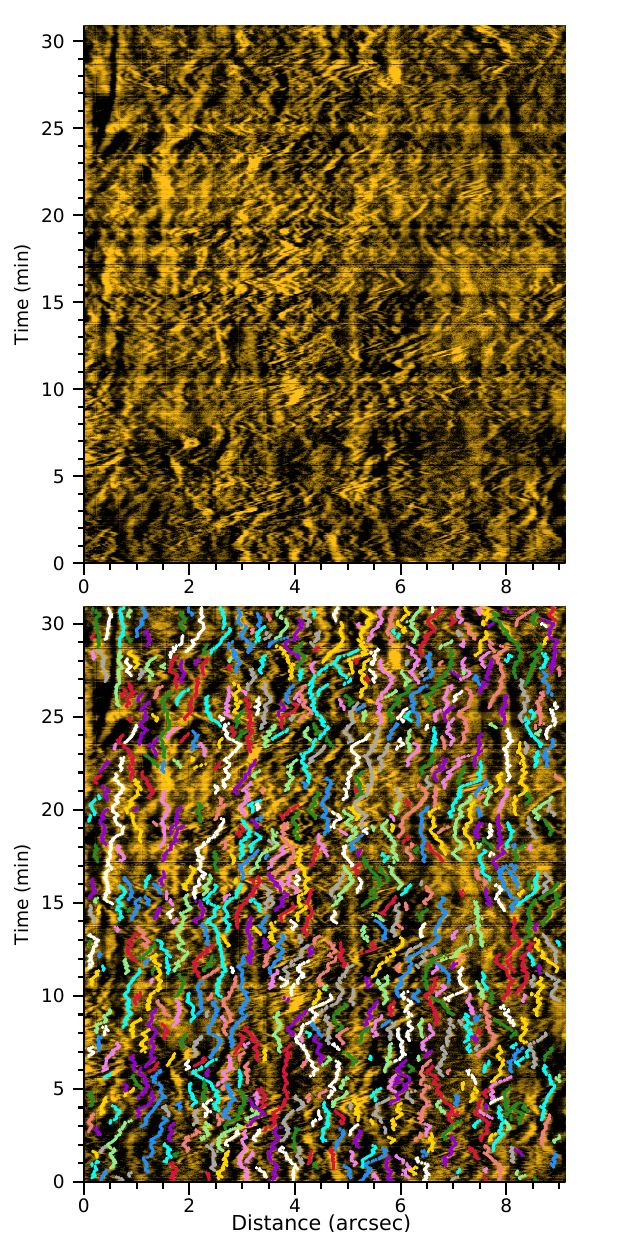}
	\caption{One example of time–distance diagrams obtained from the slits. Because the image intensities were inverted for visualization, the oscillating fibrils appear as bright thread-like structures. In the lower panel, the colored tracks indicate the reliable oscillation threads.}
	\label{fig:fig2}
\end{figure}

To extract transverse oscillations of fibrils, slits were placed in fibril-dense regions within the FOV and oriented as perpendicular as possible to the fibril axes to obtain the transverse displacements. A total of eight slit sets were defined, including six in D1 and two in D2. Although D2 contained fewer slit sets, each individual slit was substantially longer, thereby covering a larger spatial extent and sampling a large population of fibrils. Each slit set consists of three parallel slits separated by 0\farcs25, and each slit produced a time–distance diagram (an example is shown in Figure \ref{fig:fig2}).

Fibril positions and oscillation properties were extracted using the Automatic Northumbria University Wave Tracking algorithm (Auto-NUWT; \citealp{Morton+2013,Morton+2014,Weberg+2018}). The algorithm determines subpixel fibril centers through local-extrema detection and Gaussian fitting, links them into continuous motion threads using spatiotemporal neighborhood searching, and identifies the dominant oscillation periods from the detrended displacement signals using a fast Fourier transform (FFT). Displacement and velocity amplitudes are then derived from the resulting oscillatory trajectories.

To robustly identify transverse oscillations, we applied two selection criteria to the Auto-NUWT trajectories: (1) an FFT significance test ($Q_{\text{fft}}$) required the dominant spectral peak to exceed the $95\%$ confidence level; (2) a Kolmogorov–Smirnov test ($Q_{\text{KS}}$) on tracking residuals confirmed that the tracking residuals were statistically consistent with a normal distribution ($p > 0.05$). Only trajectories satisfying both criteria were accepted as reliable oscillations ($\text{Reliable} = Q_{\text{fft}} \cap Q_{\text{KS}}$).

\section{Results} \label{sec:results}

\begin{figure*}[ht!]
	\includegraphics[scale=1.3]{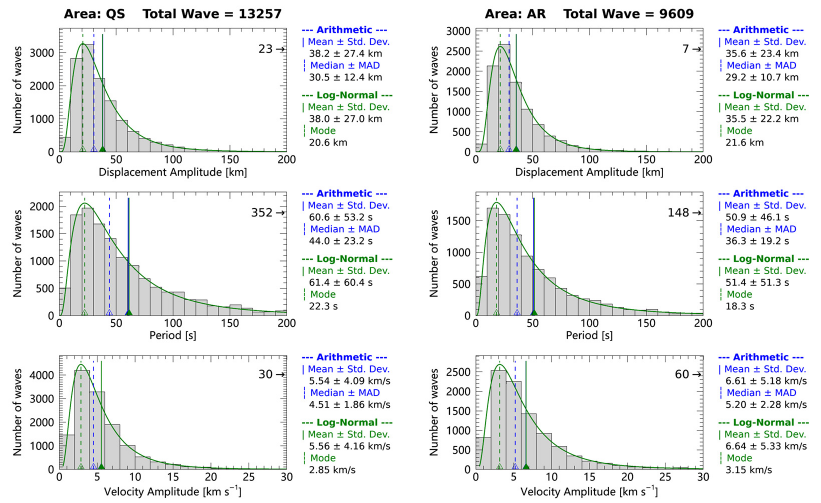}
	\caption{Histograms of wave properties for QS (left) and AR (right) fibrils. From top to bottom, the panels display the displacement amplitude, period, and velocity amplitude. Histograms denote the measured distributions, while green curves represent log-normal fits. Blue solid and dashed vertical lines mark the arithmetic mean and median, respectively, and green solid and dashed vertical lines mark the log-normal mean and mode, respectively. The total number of detected waves is indicated at the top of each column.}
	\label{fig:fig3}
\end{figure*}

A statistical analysis was performed 13,257 measurements in D1 and 9,609 measurements in D2. Figure \ref{fig:fig3} shows the statistical distributions of the displacement amplitude, period, and velocity amplitude for the two data sets. In each panel, the distribution is shown as a histogram, with a log-normal fit overplotted as a green solid curve. The number in the upper-right corner indicates the number of measurements outside the displayed range of the horizontal axis. The distributions are strongly asymmetric and positively skewed, indicating that small-amplitude and short-period oscillations constitute the majority of the detected population. This distribution pattern is also consistent with previous studies (e.g., \citealp{Jafarzadeh+2017+osc,Morton+2021RSPTA}).

The displacement amplitudes are comparable between the two regions, with values of 38.2 km in the QS and 35.6 km in the AR. The period distributions are broadly similar, although the AR distribution shows a slight shift toward shorter periods, with a mean of 50.9 s compared with 60.6 s in the QS. Consequently, the fibrils in the AR exhibit slightly larger velocity amplitudes on average. These results indicate that high-frequency waves are ubiquitous and dominant in both QS and AR regions. While the displacement amplitudes remain comparable between the two magnetic environments, fibril oscillations in the AR are characterized by higher velocity amplitudes and shorter periods on average.
 
\begin{figure*}[ht!]
	\plotone{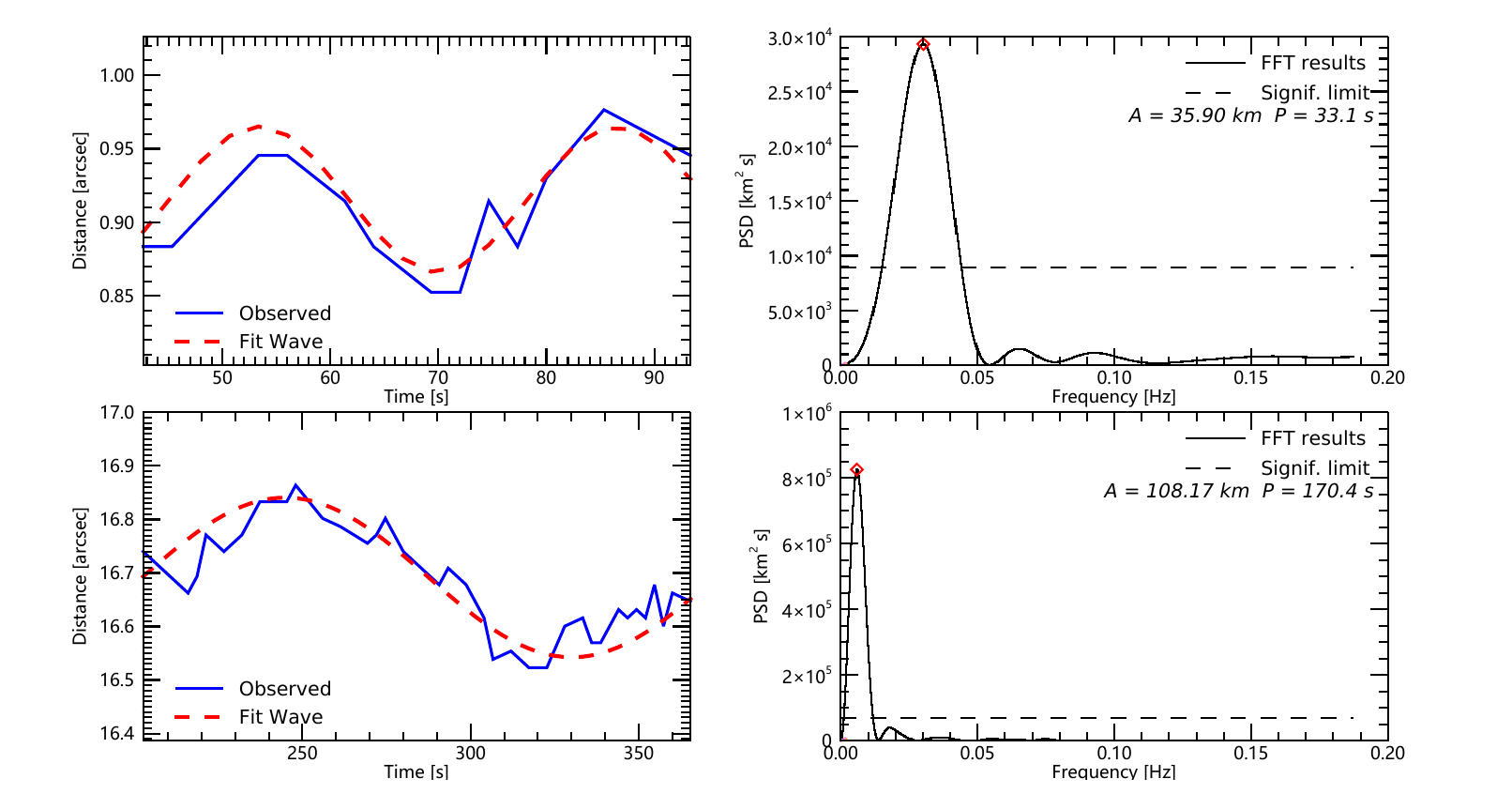}
	\caption{Examples of transverse wave signals identified in chromospheric fibrils. The upper and lower rows show two examples of a small-amplitude high-frequency oscillation and a large-amplitude low-frequency oscillation, respectively. The corresponding Fourier power spectra confirm the dominant oscillation periods, with the red diamond marking the primary peak.}
	\label{fig:fig4}
\end{figure*}

Figure \ref{fig:fig4} presents two representative examples of transverse waves identified in the fibrils. The upper panels show a small-amplitude, high-frequency wave with a displacement amplitude of 36 km and a period of 33 s. Despite its relatively small displacement, the oscillatory signal is clearly reproduced by the fit, and the corresponding Fourier power exhibits a pronounced peak above the $95\%$ confidence level (white noise background, \citealp{Torrence+1998}). Such high-frequency transverse waves have been less frequently characterized statistically in fibrils, although similar high-frequency motions have been reported in spicules and individual fibril observations \citep{Bate+2022,Petrova+2025}. In contrast, the lower panels illustrate a large-amplitude, low-frequency wave with an amplitude of 108 km and a period of 170 s. This wave is comparable to the transverse oscillations commonly reported in previous fibril studies and is likewise successfully identified by the method. These two examples demonstrate that our detection procedure is sensitive to a broad range of oscillation properties, from large-amplitude, long-period oscillations commonly reported in previous studies to small-amplitude, short-period oscillations that have been less well characterized statistically.

To measure the phase speed of transverse waves along fibrils, we compared the waves extracted from two adjacent slits. The separation between the two slits was set to $0.25^{\prime\prime}$. Since the slits were placed close to the middle positions of the fibrils, the same wave was expected to be sampled by both slits.

The wave parameters, including the dominant frequency and phase, were obtained independently for each slit using the NUWT analysis \citep{Morton+2013}. Corresponding waves between adjacent slits were identified using a matching procedure based on spatial proximity, frequency similarity, and temporal overlap. Candidate pairs were selected by requiring a mean positional separation of less than 5 pixels, a dominant-frequency difference of less than 0.001 Hz (i.e., $<5\%$ for waves with periods shorter than 50 s), and sufficient temporal overlap ($>50\%$). For the remaining candidates, normalized cross-correlation was calculated between the transverse displacement time series extracted from the overlapping portions of the two waves \citep{Tomczyk+2009}. The final correspondence was determined by minimizing a combined cost function that incorporates the spatial offset, frequency difference, and correlation coefficient using the Hungarian assignment algorithm \citep{Kuhn+1955}. This global one-to-one matching scheme was adopted to avoid ambiguous associations and duplicate pairing when multiple fibrils with similar oscillation properties were present in neighboring slits.

For each matched pair, the phase delay was derived by combining the Fourier phase difference obtained from NUWT with the time lag determined from the cross-correlation analysis. Because the Fourier phase difference is intrinsically ambiguous by integer multiples of 2$\pi$, the cross-correlation time lag was used as an independent constraint to determine the appropriate phase cycle and unwrap the phase difference. The resulting phase delay was then used to calculate the phase speed from the slit separation: $v_\text{ph} = d / \Delta t_\text{ph}$, where $d$ represents the slit separation, and $\Delta t_\text{ph}$ represents the time delay determined from the phase difference. Only wave pairs satisfying the oscillation quality criteria and yielding physically reasonable phase delays were retained. Measurements with phase speeds exceeding 100 km\,s$^{-1}$ were excluded, as they were mainly associated with uncertain wave matching or unreliable phase-delay determination, this subset accounted for $\sim10\%$ of the total.

\begin{figure}[ht!]
	\plotone{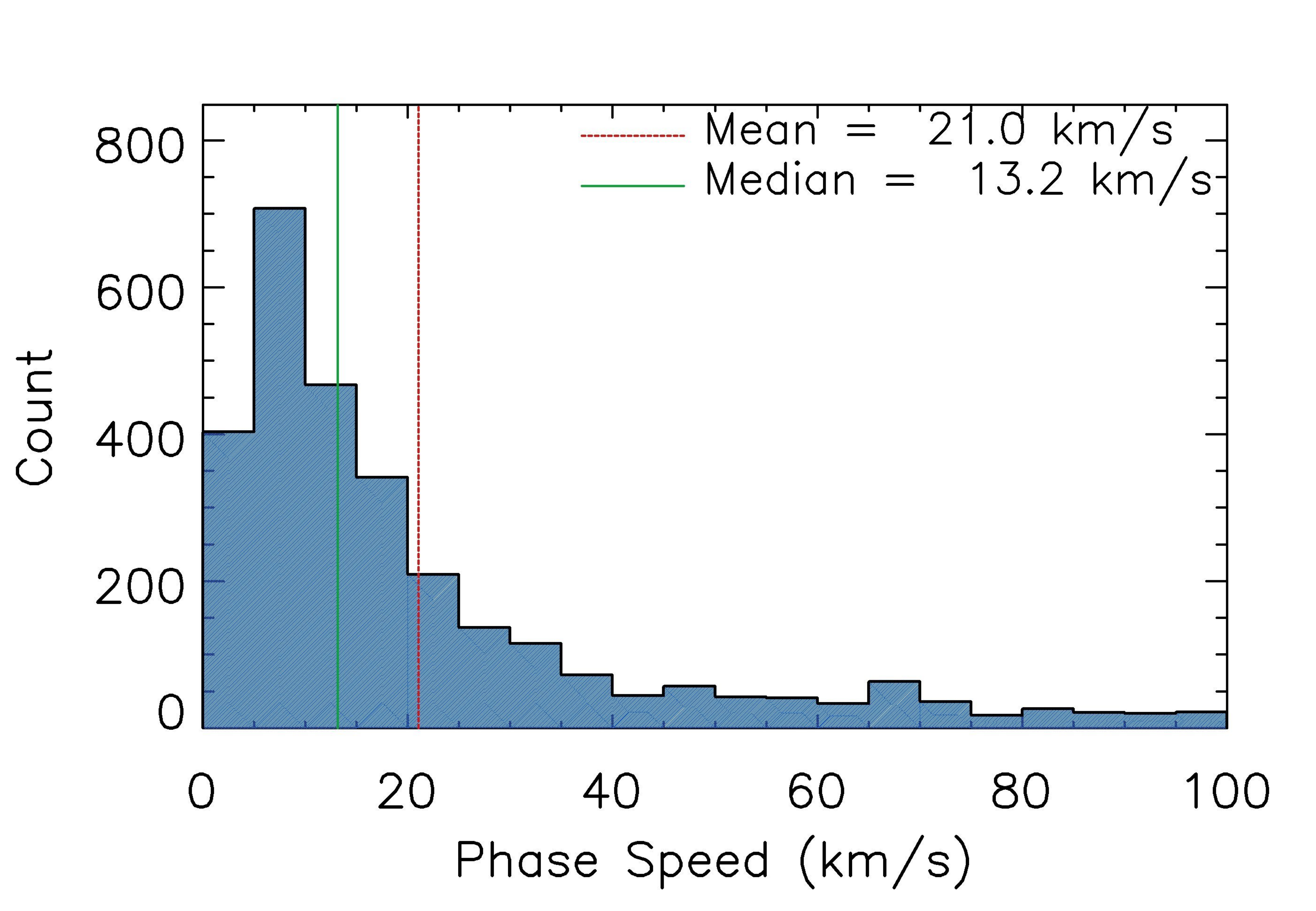}
	\caption{Histogram of the phase speeds of 3,244 propagating waves identified in the QS and AR. The green line marks the median value, and the red line marks the mean value.}
	\label{fig:fig5}
\end{figure}

Figure \ref{fig:fig5} shows the phase speed distribution of 3,244 propagating waves (approximately $14\%$ of all detected waves) that satisfied the criteria for reliable phase-speed measurements in the two regions. This fraction is comparable to those reported in previous studies (e.g., \citealp{Jafarzadeh+2017+osc,Bate+2022}). The measured phase speeds span a broad range, although the majority are below 25 km\,s$^{-1}$. The mean and median phase speeds are 21.0 km\,s$^{-1}$ and 13.2 km\,s$^{-1}$, respectively. The distribution is positively skewed owing to a small number of waves with relatively high phase speeds. The mean phase speeds of the propagating transverse waves identified between adjacent slits are 19.6 km s$^{-1}$ and 22.8 km s$^{-1}$ for the QS and AR fibrils, respectively (see Table \ref{tab:Properties}).

The observed transverse oscillations are interpreted as kink mode. Following previous studies (e.g., \citealp{Bate+2024}), the energy flux carried by these transverse waves can be estimated using the expression:
\begin{equation}
	F \approx \frac12 f \rho_i V^2 V_\text{ph}
\end{equation}
where f is the filling factor of fibrils, $\rho_i$ is the internal mass density of the flux tube, V is the transverse velocity amplitude, and $V_\text{ph}$ is the measured phase speed. The filling factor f is assumed to be $5\%$, corresponding to the upper limit estimated by \citet{Morton+2012A&A}. The internal mass density of fibrils ($\rho_i$) is adopted from the numerical simulations of \citet{Leenaarts+2012}, with values ranging from $10^{-10}$ to $10^{-9}$ kg m$^{-3}$. Using the mean transverse velocity amplitude of 6.1 km s$^{-1}$ and the mean phase speed of 21.2 km s$^{-1}$ derived from our observations, the resulting energy flux carried by the detected transverse kink waves is estimated to be $F \approx 20-200\ \mathrm{W\,m^{-2}}$. This energy is clearly insufficient to sustain chromospheric heating ($\sim 10^{4}$ W\,m$^{-2}$, \citealp{Withbroe+1977}).

\section{Discussion and Conclusion} \label{sec:dis}

Previous observations have demonstrated that transverse waves are ubiquitous in chromospheric fibrillar structures. To place the properties of the transverse oscillations detected in this work into the context of previous observations, Table \ref{tab:Comparison} summarizes the statistical properties of transverse waves reported from different chromospheric structures and magnetic environments. Direct comparisons of the measured wave properties among different studies should be made with caution, because the derived statistical distributions can be affected by differences in observational instruments, passbands, target structures, and analysis methods. For instance,\citet{Okamoto+2011ApJ} pointed out that their detection procedure was biased toward oscillations with periods shorter than 100 s, which naturally leads to a shorter characteristic period in their sample. It should be noted that most previous works, as well as this study, characterize the wave properties using mean values with standard deviations, whereas \citet{Okamoto+2011ApJ} and \citet{Kuridze+2012} provided median values with standard deviations. Nevertheless, such comparisons remain informative and provide useful context for highlighting the relative characteristics of the present results. Moreover, \citet{Morton+2014}, \citet{Mooroogen+2017}, and \citet{Bate+2022} employed NUWT-based or similar wave-tracking techniques, reducing potential biases introduced by different detection approaches.

The most notable difference between this study and previous observations is the prevalence of shorter-period and smaller-amplitude transverse oscillations detected by DKIST/VBI. The characteristic periods of fibrils derived here are approximately 50–60 s, considerably shorter than those reported in most previous H$\alpha$ observations, where typical periods generally fall within 100–200 s. High-frequency transverse waves have been reported in several earlier studies, although their statistical characterization remains limited, typically based on smaller samples or conducted in different chromospheric structures. \citet{Jafarzadeh+2017+osc} found ubiquitous transverse oscillations in slender \ion{Ca}{2} H fibrils with a characteristic period of 89 ± 29 s, while \citet{Petrova+2025} identified a 15 s transverse oscillation in an individual fibril observed with the MiHI prototype. In addition, \citet{Bate+2022} measured a mean period of approximately 54 s for transverse oscillations in spicules.

A direct comparison with \citet{Morton+2014} is particularly meaningful because both studies investigated transverse oscillations in H$\alpha$ fibrils under both QS and AR. They reported mean displacement amplitudes of 94 $\pm$ 47 km and 73 $\pm$ 36 km, with periods of 116 $\pm$ 59 s and 130 $\pm$ 92 s for QS and AR fibrils, respectively. In comparison, our DKIST/VBI observations reveal substantially smaller displacement amplitudes (38 $\pm$ 27 km and 36 $\pm$ 23 km) and shorter periods (61 $\pm$ 53 s and 51 $\pm$ 46 s) in QS and AR fibrils. Interestingly, despite the reduced displacement amplitudes and periods, the velocity amplitudes derived in this study (5.5 $\pm$ 4.1 km s$^{-1}$ and 6.6 $\pm$ 5.2 km s$^{-1}$) are comparable to those reported by \citet{Morton+2014} (5.5 $\pm$ 2.4 km s$^{-1}$ and 4.4 $\pm$ 2.4 km s$^{-1}$). This indicates that the difference is unlikely to be caused by intrinsically weaker wave motions, but instead suggests that a previously under-sampled population of short-period oscillations with small displacement amplitudes has been detected.

The phase speeds measured in this study are lower than those reported in some previous investigations of chromospheric fibrils (e.g., \citealp{Morton+2012+NC}), but are comparable to the values obtained by \citet{Jafarzadeh+2017+osc}. The considerable spread in reported phase speeds among different studies is most likely attributable to differences in the methods used to determine the phase speed. 

Using the measured transverse velocity amplitudes and phase speeds, we estimated an energy flux of 20-200 W\,m$^{-2}$ for the transverse waves under typical chromospheric conditions. This estimate is based on the mean measured phase speed. Considering the full range of observed phase speeds, the maximum energy flux can reach values of up to $\sim1000$ W\,m$^{-2}$. This value is lower than some previous estimates, such as the kink-wave energy flux of 4300 $\pm$ 2700 W\,m$^{-2}$ reported by \citet{Morton+2012+NC}. However, such differences mainly reflect uncertainties in the adopted atmospheric parameters, particularly the internal density and phase speed. For example, \citet{Jafarzadeh+2017+osc} adopted a density of $10^{-6}\ \mathrm{kg\,m^{-3}}$, which is several orders of magnitude higher than the chromospheric fibril densities adopted here, while \citet{Petrova+2025} assumed phase speeds of 50-300 km\,s$^{-1}$ for an individual fibril. Therefore, the lower energy flux obtained here does not necessarily indicate weaker wave transport, but reflects the different wave populations sampled by different observations. As noted by \citet{Verth+2016GMS}, unresolved small-amplitude or high-frequency waves may be excluded from energy estimates, leading to an incomplete assessment of the chromospheric wave-energy budget. Our results provide new constraints on this previously under-characterized high-frequency wave population.

Overall, our results demonstrate that high-frequency transverse oscillations are a common property of chromospheric fibrils in both quiet-Sun and active regions. The predominance of short-period and small-displacement waves revealed by DKIST/VBI indicates that previous observations may have sampled only a portion of the fibril wave population. By revealing a population of smaller-amplitude and higher-frequency waves in H$\alpha$ fibrils, this study provides new constraints on the role of fibrils in chromospheric wave propagation and energy transport.

\begin{deluxetable*}{lccc c ccc}
	\tabletypesize{\scriptsize}
	\tablewidth{0pt}
	\setlength{\tabcolsep}{6pt}
	\renewcommand{\arraystretch}{1}
	\tablecaption{Properties of Transverse Oscillations/Waves in Fibrils\label{tab:Properties}}
	\tablehead{
		\colhead{Quantity} &
		\multicolumn{3}{c}{Quiet Sun} &
		\colhead{} &
		\multicolumn{3}{c}{Active Region} \\
		\cline{2-4} \cline{6-8}
		\colhead{} &
		\colhead{Mean} & \colhead{Median} & \colhead{$\sigma$\tablenotemark{a}} & \colhead{}  &
		\colhead{Mean} & \colhead{Median} & \colhead{$\sigma$\tablenotemark{a}}
	}
	\startdata
	Displacement Amplitude (km)       & 38.2 & 30.5 & 27.4 &  & 35.6 & 29.2 & 23.4  \\ 
	Period (s)                        & 60.6 & 44.0 & 53.2 &  & 50.9 & 36.3 & 46.1  \\ 
	Velocity Amplitude (km\,s$^{-1}$) & 5.54 & 4.51 & 4.09 &  & 6.61 & 5.20 & 5.18  \\ 
	Phase Speed (km\,s$^{-1}$)        & 19.6 & 12.5 & 20.0 &  & 22.8 & 14.6 & 21.7  \\ 
	\enddata
	\tablenotetext{a}{Standard deviation of the distributions.}
\end{deluxetable*}

\begin{deluxetable*}{lccccccc}
	\tabletypesize{\scriptsize}
	\tablewidth{0pt}
	\setlength{\tabcolsep}{2pt}
	\renewcommand{\arraystretch}{1}
	\tablecaption{Comparison of Properties of Transverse Oscillations/Waves in Fibrils\label{tab:Comparison}}
	\tablehead{
		\colhead{Reference} & \colhead{Passband} & \colhead{Structure} & \colhead{Region} & \colhead{Displacement Amplitude} & \colhead{Period} & \colhead{Velocity Amplitude} & \colhead{Phase Speed} \\ 
		\colhead{} & \colhead{} & \colhead{} & \colhead{} & \colhead{(km)} & \colhead{(s)} & \colhead{(km\,s$^{-1}$)} & \colhead{(km\,s$^{-1}$)}
	}
	\startdata
	This study                       & H$\alpha$ & Fibril & QS & $38\pm27$ & $61\pm53$ & $5.5\pm4.1$ & $20\pm20$  \\ 
	& H$\alpha$ & Fibril & AR & $36\pm23$ & $51\pm46$ & $6.6\pm5.2$ & $23\pm22$  \\
	\citet{Okamoto+2011ApJ}       & \ion{Ca}{2} H & Spicule & CH & $55\pm50$ & $45\pm30$ & $7.4\pm3.7$ & $270\pm30$  \\ 
	\citet{Kuridze+2012}            & H$\alpha$ & Mottle & QS & $200\pm67$ & $165\pm51$ & $8.0\pm3.6$ & 40-110  \\ 
	\citet{Morton+2012+NC}             & H$\alpha$ & Fibril & QS & $315\pm130$ & $188\pm5$ & $6.4\pm2.8$ & $72\pm20$  \\
	\citet{Morton+2014}            & H$\alpha$ & Fibril & QS & $94\pm47$ & $116\pm59$ & $5.5\pm2.4$ & \nodata  \\
	& H$\alpha$ & Fibril & AR & $73\pm36$ & $130\pm92$ & $4.4\pm2.4$ & \nodata  \\ 
	\citet{Jafarzadeh+2017+osc}         & \ion{Ca}{2} H & Fibril & AR & $35\pm16$ & $89\pm29$ & $2.5\pm0.8$ & $15\pm14$  \\           
	\citet{Mooroogen+2017}          & H$\alpha$ & Fibril & AR & $128\pm43$ & $85\pm43$ & $4.2\pm1.6$ & $446\pm338$  \\  
	\citet{Bate+2022}               & H$\alpha$ & Spicule & AR & $151\pm124$ & $54\pm45$ & $21\pm13$ & $116\pm21$  \\      
	\enddata
\end{deluxetable*}

\begin{acknowledgments}
The research reported herein is based in part on data collected with the Daniel K. Inouye Solar Telescope (DKIST), a facility of the National Solar Observatory (NSO). NSO is managed by the Association of Universities for Research in Astronomy, Inc., and is funded by the National Science Foundation. Any opinions, findings and conclusions or recommendations expressed in this publication are those of the authors and do not necessarily reflect the views of the National Science Foundation or the Association of Universities for Research in Astronomy, Inc. In addition, the authors acknowledge the data support from SDO. This work is supported by the Strategic Priority Research Program of the Chinese Academy of Sciences, grant No. XDB0560000; the National Key R$\&$D Program of China, under No. 2024YFA1612001; the National Science Foundation of China (NSFC) under Nos. 12325303, 12473059, 12433010, and 12403065; Yunnan Key Laboratory of Solar Physics and Space Science under No. 202205AG070009; the Yunnan Science Foundation of China under Nos. 202501AW070002 and 202601AS070076; the Yunnan Province Xing Dian Talent Support Program, the Yunling Scholar Project, and Yunnan Revitalization Talent Support Program under Nos. 202305AS350029, 202305AT350005.
\end{acknowledgments}

\bibliography{references}{}
\bibliographystyle{aasjournalv7}

\end{document}